\documentclass[%
reprint,
superscriptaddress,
amsmath,
amssymb,
aps,
]{revtex4-2}

\usepackage{tocloft}
\usepackage{graphicx}  
\usepackage{dcolumn}   
\usepackage{bm}        
\usepackage{amssymb}   
\usepackage{amsmath}
\usepackage{float}
\usepackage{afterpage}
\usepackage{hyperref}
\usepackage{placeins}
\usepackage{chemformula}

\usepackage{mathtools}
\usepackage{graphicx}
\usepackage{dcolumn}
\usepackage{bm}
\usepackage{lipsum}
\usepackage[utf8]{inputenc}
\usepackage[T1]{fontenc}
\usepackage{mathptmx}
\usepackage{etoolbox}
\usepackage{orcidlink}
\usepackage{physics}

\usepackage{graphicx}
\usepackage{dcolumn}
\usepackage{bm}
\usepackage{multirow}
\usepackage{soul}
\usepackage[normalem]{ulem}
\usepackage{xcolor}
\usepackage{kantlipsum}
\usepackage[version=4]{mhchem}
\usepackage{textcomp}

\usepackage{booktabs}
\newcolumntype{C}{>{$}c<{$}}
\AtBeginDocument{
\heavyrulewidth=.08em
\lightrulewidth=.05em
\cmidrulewidth=.03em
\belowrulesep=.65ex
\belowbottomsep=0pt
\aboverulesep=.4ex
\abovetopsep=0pt
\cmidrulesep=\doublerulesep
\cmidrulekern=.5em
\defaultaddspace=.5em
}
\usepackage{siunitx}
\usepackage{multirow}

\usepackage{xfrac}

\usepackage{orcidlink}
\usepackage{xcolor}

\definecolor{amber}{rgb}{1,0.49,0}

\newcommand{\editor}[2]{%
  \expandafter\newcommand\csname #1note\endcsname[1]{%
    \textcolor{#2}{(\textbf{#1:} ##1)}}%
  \expandafter\newcommand\csname #1\endcsname[1]{%
    \textcolor{#2}{##1}}%
  \expandafter\newcommand\csname #1cancel\endcsname[1]{%
    \textcolor{#2}{\sout{##1}}}%
  \expandafter\newcommand\csname #1change\endcsname[2]{%
    \textcolor{#2}{\sout{##1} ##2}}%
  \newenvironment{#1text}{\color{#2}}{\color{black}}
}

\editor{ED}{purple}
\definecolor{verde}{rgb}{0.,0.6,0}
\editor{PP}{verde}
\editor{SB}{amber}
\editor{resub}{cyan}

\usepackage{mathrsfs}
\usepackage{placeins}

\begin{document}

\title{Seebeck coefficient  of ionic conductors from Bayesian regression analysis}

\author{Enrico Drigo\,\orcidlink{0000-0002-1797-2987}}
\affiliation{%
 SISSA---Scuola Internazionale Superiore di Studi Avanzati, 34136 Trieste, Italy
}
\email{endrigo@sissa.it}
\author{Stefano Baroni\,\orcidlink{0000-0002-3508-6663}}
\affiliation{%
 SISSA---Scuola Internazionale Superiore di Studi Avanzati, 34136 Trieste, Italy
}%
\affiliation{%
 CNR-IOM---Istituto Officina Materiali, DEMOCRITOS SISSA unit, 34136 Trieste, Italy
}%
\author{Paolo Pegolo\,\orcidlink{0000-0003-1491-8229}}
\affiliation{%
 SISSA---Scuola Internazionale Superiore di Studi Avanzati, 34136 Trieste, Italy
}
\affiliation{Current address: COSMO---Laboratory of Computational Science and Modeling, IMX, \'Ecole Polytechnique F\'ed\'erale de Lausanne, 1015 Lausanne, Switzerland}


\begin{abstract}
We propose a novel approach to evaluating the ionic Seebeck coefficient in electrolytes from relatively short equilibrium molecular dynamics simulations, based on the Green-Kubo theory of linear response and Bayesian regression analysis. By exploiting the probability distribution of the off-diagonal elements of a Wishart matrix, we develop a consistent and unbiased estimator for the Seebeck coefficient, whose statistical uncertainty can be arbitrarily reduced in the long-time limit. We assess the efficacy of our method by benchmarking it against extensive equilibrium molecular dynamics simulations conducted on molten $\ce{CsF}$ using empirical force fields. We then employ this procedure to calculate the Seebeck coefficient of molten $\ce{NaCl}$, $\ce{KCl}$ and $\ce{LiCl}$ using neural network force fields trained on \textit{ab initio} data over a range of pressure-temperature conditions.
\end{abstract}

\maketitle

\section{Introduction}

The out-of-equilibrium response of an extended system to an external perturbation is the object of transport theory~\cite{degroot2004}. In 1931, Onsager presented a theory for the response of macroscopic fluxes of locally conserved quantities to thermodynamic forces, i.e., to the gradients of the intensive variables conjugate to the quantity being transported~\cite{onsager1931reciprocal1,onsager1931reciprocal2}. Onsager's theory, in conjunction with the seminal contributions of Green and Kubo during the 1950s~\cite{green1952, green1954,kubo1957a,kubo1957b}, forms the cornerstone of contemporary transport theory. Despite its significance, certain misconceptions and numerical difficulties, only recently overcome, led to the misguided notion that an equilibrium-based approach to calculate the heat conductivity would not be feasible in principle within a fully \emph{ab initio} approach~\cite{stackhouse2010}, while the calculation of other transport coefficients, such as, for example, electric conductivity or viscosity, would be extremely numerically demanding even using classical force fields. A recent resurgence of interest has sought to clarify and correct these misconceptions within transport theory, leading to the concept of \emph{gauge invariance of transport coefficients.}~\cite{baroni2020heat,bertossa2019theory,marcolongo2016microscopic,grasselli2021invariance}.

Amid transport phenomena, the thermoelectric effect is of paramount importance in energy storage and conversion, as well as other technological applications~\cite{Snyder2008,biswas2012high,zhu2017compromise,chu2017path,wang2019flexible,yu2020thermosensitive}. As early as 1794, A. Volta observed that a temperature gradient could induce an electromotive force under open circuit conditions~\cite{volta1794}, subsequently termed the \textit{Seebeck effect} after T. J. Seebeck, who independently rediscovered it in 1821~\cite{seebeck1826}. In the context of liquid electrolytes, a thermal gradient induces an electric current~\cite{galamba2007equilibrium,grisafi2023, kjelstrup2015, bedeaux2023}, while liquid insulators manifest macroscopic electric polarization~\cite{drigo2023seebeck}.

Among the techniques available for computing transport coefficients, nonequilibrium molecular dynamics (MD) is one of the most commonly used~\cite{salanne2011thermal,bonella2017thermal, ohtori2009calculations,wirnsberger2016,wirnsberger2017,Bresme2008,dilecce2018,DiLecce2017}. Nonequilibrium MD simulations mimic the presence of a thermodynamic driving force by introducing the explicit representation of an external perturbation in the dynamics of the system. In the case of temperature gradients, for example, this involves the introduction of a pair of heat source/sink separated in space in the simulation cell. Careful consideration is necessary to mitigate nonlinear effects in these simulations, whose statistical precision is also difficult to assess~\cite{alexis2009}. In contrast, the GK formulation of  transport theory exclusively draws upon equilibrium MD simulations, thus in principle simplifying the procedure and facilitating the statistical analysis of the results. Despite this advantage, the GK estimator is beset by numerical noise, mandating the use of extremely long equilibrium MD  trajectories~\cite{baroni2020heat}. In recent years, a specific statistical technique called \textit{cepstral analysis} has emerged, which provides a drastic reduction in the computational burden associated with GK calculations, making the procedure affordable~\cite{ercole2017accurate}. As a drawback, cepstral analysis is only applicable to diagonal Onsager coefficients, precluding its direct application to thermoelectric phenomena. The calculation of the Seebeck coefficient is further complicated by the necessity to compute molecular partial enthalpies, which enter the definition of the heat current\cite{degroot2004} and can be obtained through the fluctuations method proposed by Debenedetti~\cite{debenedetti1987,debenedetti1987_1,debenedetti1988_2}. 

The primary focus of this study is the efficient computation of the Seebeck coefficient in electrolytes through the GK method. Leveraging  gauge invariance of transport coefficients and cepstral analysis, we develop a Bayesian regression framework based on the statistical properties of the GK estimator for off-diagonal Onsager coefficients. We test our method by comparing it with traditional GK calculations done on extensive equilibrium MD  simulations of common molten alkali halides. In particular, we use as benchmark calculations a $\ce{CsF}$ model parameterized by the inexpensive Born-Mayer-Huggins-Tosi-Fumi rigid ion inter-ionic potential~\cite{FUMI1964_1,TOSI1964_2}, and we then apply our Bayesian approach to compute the Seebeck coefficient of $\ce{NaCl}$, $\ce{KCl}$ and $\ce{LiCl}$ systems modeled through \textit{ab initio}-accurate state-of-the-art neural network potentials.

The structure of the article is as follows. Sec.~\ref{sec:theory} contains reviews of Onsager's theory of thermoelectricity and cepstral analysis, as well as an extension of the latter to off-diagonal Onsager coefficients; our Bayesian regression scheme concludes this section. 
In Sec.~\ref{sec:numerical} we present the results of numerical experiments to both validate and apply the Bayesian regression procedure to compute the Seebeck coefficient of common molten salts. 
Finally, Supporting Information Sec. A contains a proof of the gauge invariance of the off-diagonal Onsager coefficients, and Supporting Information Sec. B demonstrates an application of the Bayesian approach to diagonal transport coefficients, for which conventional cepstral analysis can already be applied.

\section{Theory}\label{sec:theory}

\subsection{Green-Kubo theory of thermoelectricity}\label{ssec:gk}

Linear response theory assumes that the macroscopic flux of a locally conserved density, such as charge and energy, is linearly related to the gradients of the intensive thermodynamic variable conjugated to the conserved ones. In the context of thermoelectric systems, the Onsager equations provide expressions for the charge and heat fluxes\cite{onsager1931reciprocal1, onsager1931reciprocal2, baroni2020heat}

\begin{align}\label{eq:Onsager}
    \begin{split}
        \mathbf{J}_c = \sigma \mathbf{E} - K_{12} \frac{\grad T}{T}, \\
        \mathbf{J}_q = K_{12}e \mathbf{E} - L_{qq} \frac{\grad T}{T},
    \end{split}
\end{align}
where $V$ is the volume of the system, $T$ its temperature, $\mathbf{J}_c$ is the electric flux, $\mathbf{J}_q = \mathbf{J}_e - \sum_i h_i\mathbf{J}_i $ the heat flux, $\mathbf{J}_e$ the energy flux, $\mathbf{J}_i$ the number flux related to the $i$th molecular species, $h_i$ the molecular entalpy, and $\mathbf{E}$ the applied electric field; the ionic conductivity, $\sigma$, and the quantities $K_{12}$ and $L_{qq}$ are proportional to some Onsager coefficients.

The GK theory provides a microscopic formulation of the Onsager coefficients, expressed through the time-integral of the correlation function of the associated fluxes~\cite{green1952,green1954,kubo1957a,kubo1957b,kadanoff1963hydrodynamic}:

\begin{align}
    \sigma = \frac{V}{3 k_B T} \int_0^\infty \left\langle\bm{J}_c(t) \cdot \bm{J}_c(0) \right\rangle \dd{t}, \label{eq: GKsigma} \\
    K_{12} = \frac{V}{3 k_B T^2} \int_0^\infty \left\langle\bm{J}_c(t) \cdot \bm{J}_q(0) \right\rangle \dd{t}. \label{eq: GKK_12}\\
    L_{qq} = \frac{V}{3 k_B T} \int_0^\infty \left\langle\bm{J}_q(t) \cdot \bm{J}_q(0) \right\rangle \dd{t}. \label{eq: GKLqq} 
\end{align}
Here, angled brackets denote the equilibrium average over the initial conditions of an equilibrium MD  trajectory, and the italic font, as in $\bm{J}$, indicates an implicit dependence of the quantity on phase-space variables. To compute the heat flux, we need to determine the partial molecular enthalpies which we do using the fluctuation method proposed by Debenedetti~\cite{debenedetti1987, debenedetti1987_1, debenedetti1988_2}. This method enables us to compute the molecular enthalpies from the same microcanonical trajectory used to sample the energy and the charge fluxes.
By analyzing the Onsager equations, the Seebeck coefficient, denoted as $S$, can be calculated from Eq.~\eqref{eq:Onsager} as

\begin{equation}\label{eq:seebeckonsager}
	S = \left.\frac{{E}}{\grad T }\right\vert_{\mathbf{J}_c =0}=\frac{K_{12}}{\sigma T}, 
\end{equation}
where $E$ and $\grad T$ represent any Cartesian components of the electric field and temperature gradient, respectively.

\subsection{Estimation of transport coefficients}\label{ssec:estimation}

To achieve statistically significant results, the evaluation of GK integrals such as those appearing in Eqs.~\eqref{eq: GKsigma}, \eqref{eq: GKK_12}, and \eqref{eq: GKLqq} needs long equilibrium MD simulations to gather enough data to accurately capture the correlation functions' tails~\cite{ercole2017accurate,grasselli2021invariance}. Even when this is done, determining the ideal number of blocks for computing error bars on the Onsager coefficients remains challenging and can lead to varying outcomes~\cite{baroni2020heat}. As mentioned above, cepstral analysis~\cite{ercole2017accurate, bertossa2019theory} proved valuable in addressing this matter. This method draws upon the Wiener-Khinchin theorem~\cite{wiener1930generalized, khintchine1934korrelationstheorie}, which states that GK integrals can be expressed as the zero-frequency values of power spectral densities, i.e.,

\begin{align}\label{eq:psd}
    \mathcal{S}^{ij}(\omega) = \int_{0}^{\infty} \expval{\widehat{\mathbf{J}}^i(t)\cdot \widehat{\mathbf{J}}^j(0)} e^{i \omega t} \dd{t}.
\end{align}
Here, hats over variable symbols indicate samples drawn from a stochastic process.
The problem of computing GK integrals is thus recast into the evaluation of the zero-frequency limit of the power spectrum~\cite{baroni2020heat, grasselli2021invariance}.
Cepstral analysis, commonly used in speech recognition~\cite{bogert1963quefrency}, effectively filters out the noise from a signal's power spectrum by leveraging its smoothness and statistical properties~\cite{ercole2017accurate, bertossa2019theory}. 
However, it is essential to acknowledge that this approach is solely applicable to diagonal Onsager coefficients, such as ionic and thermal conductivities. In essence, for diagonal Onsager coefficients, the estimator of the power spectrum---termed the \emph{periodogram}, ${\widehat{\mathcal{S}}}^{ii}_k$---at a given discrete frequency, $\omega_k=\frac{2\pi k}{\tau}$, with $\tau$ denoting the length of the equilibrium MD  trajectory, is distributed as the true power spectrum, $\mathcal{S}^{ii}(\omega_k)$, multiplied by a normalized chi-square random variable with $\nu$ degrees of freedom~\cite{ercole2017accurate,bertossa2019theory}:

\begin{align}\label{eq:periodogram}
    \widehat{\mathcal{S}}^{ii}_k = \mathcal{S}^{ii}(\omega_k) \widehat{\xi}_k, \qquad \widehat{\xi}_k \sim \frac{1}{\nu} \chi^2_\nu,
\end{align}
where ``$\sim$'' means ``is distributed as'', and $\nu=2\ell$, with $\ell$ the number of independent flux samples (in isotropic media $\ell=3$, the number of Cartesian directions). However, this estimator lacks consistency because its variance is independent of the length of the equilibrium MD trajectory~\cite{ercole2017accurate}. 
The multiplicative noise, $\widehat{\xi}_k$, can be dealt with recognizing that $\widehat{\mathcal{S}}^{ii}_k$ is always positive.
Thus, applying the logarithm function to both sides of the equation leads to

\begin{align}\label{eq:log periodogram prequel}
    \log \widehat{\mathcal{S}}^{ii}_k = \log \mathcal{S}^{ii}(\omega_k) + \log \widehat{\xi}_k.
\end{align}
Since the noise on the logarithm of the power spectrum is additive, its magnitude can be reduced by a low-pass filter, making the estimator accurate and consistent.
Several studies in the literature have illustrated the successful application of cepstral analysis across diverse practical situations, ranging from heat\cite{ercole2017accurate, baroni2020heat, bertossa2019theory, grasselli2020heat, marcolongo2020gauge, tisi2021heat, lundgren2021mode, pegolo2022temperature, pegolo2024thermal}, charge\cite{pegolo2020oxidation, rozsa2021solvation, pegolo2022topology,pegolo2023self}, and momentum\cite{malosso2022viscosity} transport.
The challenge lies in the case of off-diagonal coefficients, since $\widehat{\mathcal{S}}^{cq}_k$ can assume negative values, rendering the aforementioned procedure impossible. Nevertheless, exploiting the statistical properties of off-diagonal periodograms remains a viable option. The real part of $\widehat{\mathcal{S}}^{cq}_k$ is in fact distributed according to the Variance-Gamma distribution~\cite{pearson1929distribution,peddada1991proof}, whose probability density function reads~\cite{nestler2019variance}

\begin{widetext}
\begin{align}\label{eq: seebeck distribution}
    \begin{multlined}
        p\left(\widehat{\mathcal{S}}^{cq}_k\right) = \frac{\left|\widehat{\mathcal{S}}^{cq}_k\right|^{\frac{\nu-1}{2}}}{\Gamma\left(\frac{\nu}{2}\right)\sqrt{2^{\nu-1}\pi(1-\rho(\omega_k)^2)}[\sigma(\omega_k) L^{qq}(\omega_k)]^{\frac{\nu+1}{4}}} \\
        \times \mathcal{K}_{\frac{\nu-1}{2}}\left(\frac{\left|\widehat{\mathcal{S}}^{cq}_k\right|}{[\sigma(\omega_k) L^{qq}(\omega_k)]^{1/2}(1-\rho(\omega_k)^2)}\right)\exp{\frac{\rho(\omega_k)\widehat{\mathcal{S}}^{cq}_k}{[\sigma(\omega_k) L^{qq}(\omega_k)]^{1/2}(1-\rho(\omega_k)^2)}},
    \end{multlined}
\end{align}
\end{widetext}
where $\Gamma$ is the Euler gamma function~\cite{EulerGamma}, $\mathcal{K}$ is the modified Bessel function of second kind~\cite{BesselSecond}, and $\rho(\omega_k)$ is the correlation coefficient, bounded by $-1< \rho < 1$. The correlation coefficient estimator is defined as 

\begin{align}\label{eq: correlation coefficient}
    \widehat{\rho}_k=\widehat{\mathcal{S}}^{cq}_k(\sigma_k L^{qq}_k)^{-1/2},
\end{align}
where ${\sigma}_k$ and ${L}^{qq}_k$ are the power spectra of the charge and heat fluxes, respectively. The probability density function of the correlation coefficient estimator is obtained from Eq.~\eqref{eq: seebeck distribution} as

\begin{align}\label{eq:likelihood}
    \begin{multlined}
        p\left(\widehat{\rho}_k\right) \propto \frac{\left|\widehat{\rho}_k\right|^{\frac{\nu-1}{2}}}{\sqrt{1-\rho(\omega_k)^2}}\mathcal{K}_{\frac{\nu-1}{2}}\left(\frac{\left|\widehat{\rho}_k\right|}{1-\rho(\omega_k)^2}\right)\\
        \times \exp{\frac{\rho(\omega_k)\widehat{\rho}_k}{1-\rho(\omega_k)^2}},    
    \end{multlined}
\end{align}
where each $\widehat{\rho}_k$ is independent of the others. Once the distribution of the samples is known, it becomes feasible to accurately estimate the correlation coefficients, and thus determine the off-diagonal Onsager coefficient and the Seebeck coefficient.

\subsection{Bayesian regression of off-diagonal Onsager coefficients}\label{ssec:bayesian}

Through Bayesian regression, we can estimate the parameters entering the probability distribution of the correlation coefficient. This ultimately allows us to access both the expectation value and the uncertainty on the Seebeck coefficient. We assume that the spectral representation of the correlation coefficient, Eq.~\eqref{eq: correlation coefficient}, can be represented by a cubic spline model \cite{de1978practical}, $\mathfrak{S}_c$,

\begin{align}\label{eq:model}
    \rho(\omega_k) \approx \mathfrak{S}_c\left(\omega_k\vert \bm{\theta}\right)=\begin{cases}
    C_1(\omega_k),\ \omega^0=0\le\omega_k\le\omega^1\\
    \vdots\\
    C_i(\omega_k),\ \omega^{i-1}\le\omega_k\le\omega^i\\
    \vdots\\
    C_{P}(\omega_k),\ \omega^{P-1}\le\omega_k\le\omega^P
    \end{cases}
\end{align}
where $C_i(\omega)=a_i + b_i \omega +c_i\omega^2 + d_i\omega^3$, with $C_i(\omega^{i-1})\doteq\theta_{i-1}$; $\mathfrak{S}_c\left(\omega_k\vert \bm{\theta}\right)$ is twice-differentiable with respect to its argument, $\omega_k$; ${\bm{\theta}=\{\theta_0,\dots, \theta_{P-1}\}}$ represent the $P$ parameters of the model optimized through Bayesian regression; the frequencies $\omega^0,\ldots,\omega^{P-1}$ are the (fixed) values at which the spline is evaluated, usually called \emph{knots}. Note the superscript labels that distinguish them from the sample frequencies. The optimization process requires sampling from the posterior parameter distribution given the dataset, ${\mathcal{D} \equiv \{\widehat{\rho}_k\}}$, i.e.

\begin{align}
    p(\widehat{\bm{\theta}}|\mathcal{D}) = \frac{p(\widehat{\mathcal{D}}|\bm{\theta}) p(\widehat{\bm{\theta}})}{p(\widehat{\mathcal{D}})},
\end{align}
where $p(\widehat{\mathcal{D}}\vert \bm{\theta})=\prod_k p\left(\widehat{\rho}_k\right)$ is the Likelihood distribution function and $p(\widehat{\bm{\theta}})$ and $p(\widehat{\mathcal{D}})$ are the prior distributions of the parameters and the dataset, respectively. The probability distribution of the dataset under a specific model choice, $p(\widehat{\mathcal{D}}|\bm{\theta})$, derives from Eq.~\eqref{eq:likelihood}, assuming the reference model is $\rho(\omega_k)=\mathfrak{S}_c\left(\omega_k\vert \bm{\theta}\right)$, as specified in Eq.~\eqref{eq:model}. The optimal parameters are estimated as the expectation values of the parameters drawn from the posterior distribution, 
 
\begin{align}\label{eq:integral}
\mathbb{E}(\widehat{\bm{\theta}}) = \int\bm{\theta} p(\widehat{\bm{\theta}}\vert \mathcal{D})\, \dd\bm{\theta}.
\end{align}
Since the integral in Eq.~\eqref{eq:integral} is analytically intractable, we rely on a Markov chain Monte Carlo (MCMC) approach to sample from the posterior probability distribution $\mathcal{N}$ realizations of the parameters, ${\widehat{\bm{\theta}}\sim p(\widehat{\bm{\theta}}\vert \mathcal{D})}$. These realizations are then averaged to estimate the integral. According to Eq.~\eqref{eq:seebeckonsager}, the Seebeck coefficient is computed as 
 
\begin{align}\label{eq: seebeck mcmc}
    S = \frac{1}{T} \mathbb{E}(\widehat{\theta}_0)\sqrt{\frac{L^{qq}_0}{\sigma_0}}.    
\end{align}
The statistical uncertainty on the Seebeck coefficient is evaluated as the standard deviation of the marginalized distribution of $\widehat{\theta}_0$. To determine the optimal number of parameters, one can adopt various model selection methods. In our implementation, we utilize the Akaike information criterion (AIC)~\cite{akaike1974new}, which tries to balance the model's accuracy and simplicity. For a model with $P$ parameters, the AIC statistics is given by

\begin{equation}
    \mathrm{AIC}(P)=-2\max_{\bm{\theta}}\log\left[p(\widehat{\mathcal{D}}\vert \bm{\theta}, P)\right] + 2P.
\end{equation}
The model with $P^\star$ parameters minimizing the AIC is regarded as the optimal choice, where ${P^\star=\operatorname{arg\,min}_{P} \mathrm{AIC}(P)}$. For the numerical implementation of the MCMC, we rely on the \texttt{emcee} code~\cite{emcee,goodman2010}. The implementation of the Bayesian regression procedure for the evaluation of the off-diagonal Onsager coefficients is available as a patch to the \textsc{SporTran} code~\cite{ercole2022sportran} at GitHub: \url{https://github.com/ppegolo/sportran/tree/bayes}, and it will be part of a future release.

\section{Numerical experiments}\label{sec:numerical}

\subsection{Markov chain Monte Carlo calculations}\label{ssec:mcmc}

\begin{figure}
    \centering
    \includegraphics[width=\linewidth]{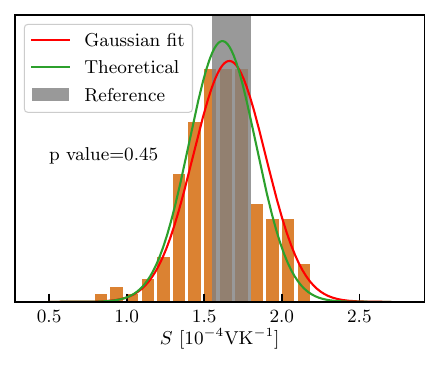}
    \caption{Distribution of the Bayesian estimate of the Seebeck coefficient extracted from 200 $2\,\mathrm{ns}$-long trajectories of \ce{CsF} at $1200\,\mathrm{K}$ and $0\,\mathrm{bar}$, and the p-value of the Shapiro-Wilk normality test. The solid red line is a Gaussian curve fitted on the distribution of the Bayesian predictions, the solid green line is a Gaussian curve centered on the average of the predictions with, as standard deviation, the average of the uncertainty estimated from the Bayesian analysis. The vertical gray band represent the reference GK estimate of the Seebeck coefficient estimated via block analysis from a $40\,\mathrm{ns}$-long simulation. }
    \label{fig:gaussianity}
\end{figure}

Given a number $P$ of parameters in the model of Eq.~\eqref{eq:model}, we define $P$ equispaced frequencies $\omega^0=0, \ldots,\omega^{P-1}=f^\star$, with $f^\star$ the periodogram's Nyquist frequency. These frequencies are the spline's knots where the fixed points $C(\omega^i)=\theta_i$, i.e., the parameters to be optimized along the MCMC, are evaluated. The initial state of each MCMC is defined to be the moving average~\cite{MovingAverage} of the periodogram at the spline knots plus a realization of Gaussian random noise.

We sample the parameters $\widehat{\bm{\theta}}$ according to the posterior distribution function $p(\widehat{\bm{\theta}}\vert \mathcal{D})$~\cite{emcee} with a uniform prior. The integral in Eq.~\eqref{eq:integral} is evaluated as the average of the generated samples, ${\mathbb{E}(\widehat{\bm{\theta}}) \simeq \overline{\bm{\theta}} = \frac{1}{\mathcal{N}}\sum_{i=1}^{\mathcal{N}}\widehat{\bm{\theta}}_i}$. To ensure reliable convergence, we compute the MCMC autocorrelation time, $\tau_\mathrm{MCMC}$, and we use it to stop the chain at the $N$th iteration, such that $N \ge 100 \tau_\mathrm{MCMC}$. Notably, the \texttt{emcee} implementation employs the \emph{stretch move} algorithm, substantially reducing the autocorrelation time compared to the Metropolis-Hastings algorithm. Details about the stretch move can be found in Refs.~\cite{emcee, goodman2010}. 

The optimal number of parameters is selected via the AIC, considering models encompassing 3 to 18 parameters. It should be noted that we enforce the (even) parity of the power spectrum by computing the spline on a mirrored version of the periodogram at zero frequency, so that a model with $P$ parameters actually corresponds to a $(2P-1)$-node cubic spline. We find that the optimal number of parameters never exceeds 16. Due to the functional form of the model in Eq.~\eqref{eq:model}, the final Seebeck coefficient estimate is $S=T^{-1}\overline{\theta}_0(\overline{L}^{qq}_0/\overline{\sigma}_0)^{1/2}$, where $\overline{L}^{qq}_0$, $\overline{\sigma}_0$ are the zero-frequency limits of the charge-charge and heat-heat power spectrum evaluated via cepstral analysis, respectively, while $\overline{\theta}_0$ represents the value of spline model at zero frequency, optimized through the MCMC. 

\subsection{MD simulations}\label{ssec:md}

We computed the Seebeck coefficient of the molten alkali halides \ce{CsF}, \ce{NaCl}, \ce{KCl}, and \ce{LiCl}, through the procedure outlined above, and benchmarked the results of \ce{CsF} against the GK ones. All the equilibrium MD  simulations are performed in \texttt{LAMMPS}~\cite{LAMMPS}. 

\subsubsection{Benchmark}

\begin{figure}
    \centering
    \includegraphics[width=\linewidth]{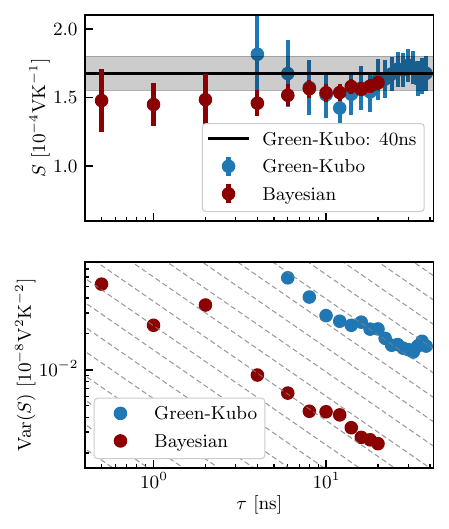} \caption{Upper panel: Seebeck coefficient of $\ce{CsF}$ at $1200\,\mathrm{K}$ and $0\,\mathrm{bar}$ computed from GK and the Bayesian regression as a function of the length of the equilibrium MD  trajectory.    Lower panel: Variance of the Seebeck coefficient of $\ce{CsF}$ at $1200\,\mathrm{K}$ and $0 \,\mathrm{bar}$ computed from GK and the Bayesian regression as a function of the length of the equilibrium MD  trajectory. The dashed line indicates the $\tau^{-1}$ scaling.}
    \label{fig:convergence_var_GK}
\end{figure}


\begin{figure*}
\centering
    \begin{minipage}{0.49\linewidth}
    \centering
    \includegraphics[width=\linewidth]{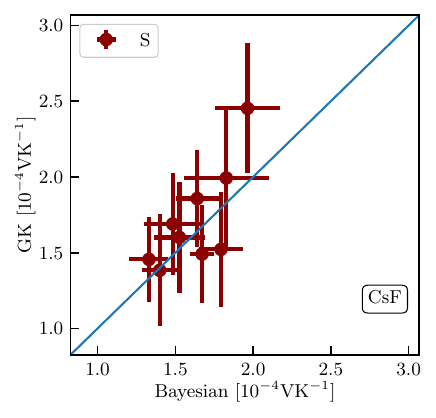}
    \end{minipage}\hfill
    \begin{minipage}{0.49\linewidth}
    \centering
    \begin{tabular}{cccc}
    \toprule
      $p$  & $T$  &  $S_{\mathrm{GK}}$   & $S_{\mathrm{Bayes}}$   \\ 
       $[\mathrm{kbar}]$ & $[\mathrm{K}]$ &  $[10^{-4}\mathrm{VK^{-1}}]$  &$[10^{-4}\mathrm{VK^{-1}}]$  \\ 
      \midrule
      $0$ & $1000$  & $1.60\pm 0.36$ & $1.52 \pm 0.17$ \\
       $0$ & $1200$  & $1.69\pm 0.34$ & $1.48\pm 0.19$ \\
       $0$ & $1400$  & $1.46\pm 0.28$ & $1.33\pm 0.13$ \\
       $1$ & $1000$  & $1.99\pm 0.47$ & $1.83\pm 0.27$ \\
       $1$ & $1200$  & $1.86\pm 0.32$ & $1.64\pm 0.14$ \\
       $1$ & $1400$  & $1.38\pm 0.37$ & $1.40\pm 0.12$ \\
       $2$ & $1000$  & $2.45\pm 0.43$ & $1.96\pm 0.21$ \\
       $2$ & $1200$  & $1.52\pm 0.38$ & $1.79\pm 0.14$ \\
       $2$ & $1400$  & $1.49\pm 0.32$ & $1.67\pm 0.08$ \\ 
       \bottomrule
    \end{tabular}
    \vspace{0.7cm}
    \end{minipage}
    \caption{Comparison between the Seebeck coefficient of molten $\ce{CsF}$ at different pressure and temperature conditions obtained through Bayesian regression from a $2\,\mathrm{ns}$-long equilibrium MD  trajectory, and from the reference GK estimate, computed via block analysis on $40\,\mathrm{ns}$-long equilibrium MD  trajectory.}
    \label{fig:comparison_materials}
\end{figure*}

The \ce{CsF} system is modeled as a sample of $432$ atoms interacting through the Born-Mayer-Huggins-Tosi-Fumi force field~\cite{FUMI1964_1, TOSI1964_2}. The parameters of the force field are taken from Ref.~\cite{wang2014}. In order to study different temperature and pressure conditions, we equilibrate the systems in the $NpT$ ensemble for $200\,\mathrm{ps}$ and in the $NVT$ ensemble for $200 \,\mathrm{ps}$. Production data for benchmarking purposes is harvested from $NVE$ trajectories whose length range from $40$ to $400\,\mathrm{ns}$. Energy and charge fluxes are sampled every fs. We use as reference the GK result obtained averaging over 40 segments of $1\,\mathrm{ns}$.

First, we test whether the Bayesian estimate is statistically reliable. To do so, we compute the Seebeck coefficient over 200 $2\,\mathrm{ns}$-long trajectories of \ce{CsF} at $1200\,\mathrm{K}$ and $0\,\mathrm{bar}$, and we compute the histogram of such values, which are shown in Fig.~\ref{fig:gaussianity}. The (red) Gaussian curve fitted on the distribution of the Seebeck coefficients is in fair agreement with the theoretical distribution (in green), whose mean and variance are estimated averaging the Bayesian prediction and uncertainty of each segment. The distribution of the Seebeck coefficients thus computed passes the Shapiro-Wilk normality test~\cite{malosso2022viscosity,shapirowilk} at a standard significance level of $0.05$. It is essential to note that trajectories of shorter duration do not withstand the normality test at the same level of significance. Based on this analysis, we conclude that $2\,\mathrm{ns}$-long trajectories are sufficient for reliably estimating the Seebeck coefficient via the Bayesian method in the context of equilibrium MD  simulations, as opposed to the more demanding GK approach.  These findings suggest that the computed Seebeck coefficient is robust with respect to the cepstral estimates of the diagonal transport coefficients entering Eq.~\eqref{eq: seebeck mcmc}, which are independently obtained before each Bayesian analysis. Consequently, production runs are carried out for $2\,\mathrm{ns}$.

As highlighted in our theoretical discussion, we emphasized the lack of consistency in the Green-Kubo (GK) estimator due to its variance's independence on the length of the Equilibrium Molecular Dynamics (equilibrium MD ) trajectory. This consistency is solely reestablished through computationally intensive block analysis. Regrettably, this approach is computationally demanding. We showcase the efficiency of the Bayesian estimator in Fig.~\ref{fig:convergence_var_GK}, where we compare the GK estimator and the Bayesian one. In the case of the former, the trajectory is divided into $1\,\mathrm{ns}$-long segments to perform block averages. In the upper panel of Fig.~\ref{fig:convergence_var_GK}, we display the GK estimates and the Bayesian regression ones as functions of the length of the equilibrium MD  trajectory. The Bayesian result is compatible with the reference one even with trajectories as short as $500\,\mathrm{ps}$, while GK calculations require at least $5\,\mathrm{ns}$ to achieve compatibility. In the lower panel of Fig.~\ref{fig:convergence_var_GK}, we compare the estimated variance of the Seebeck coefficient given by the two methods. The variance of the Bayesian results is approximately one order of magnitude lower than the GK one, confirming the higher statistical accuracy of the former.
The scaling behavior of the variance associated with the Bayesian estimate follows a $\tau^{-1}$ pattern, elucidating the reduction in variance as the length of the trajectory increases. This outcome convincingly establishes that the variance associated with the Bayesian regression approach decreases proportionally with the length of the trajectory, thus making the Bayesian estimator consistent.

As a final test, we performed simulations on molten $\ce{CsF}$ at different pressures and temperatures to compare the Seebeck coefficient computed with the Bayesian method on $2\,\mathrm{ns}$-long trajectories and the GK results on $40\,\mathrm{ns}$-long trajectories. The results are presented in Fig.~\ref{fig:comparison_materials}, revealing a satisfactory agreement between the Bayesian regression and the GK method under the thermodynamic conditions considered. The largest discrepancy between the two methods occurs near the melting point, as reported in Fig. \ref{fig:comparison_materials} at $2\,\mathrm{kbar}$ and $1000\,\mathrm{K}$. Specifically, the vanishing ionic diffusion in proximity of the solid phase causes both the numerator and denominator of Eq.~\eqref{eq:seebeckonsager} to approach zero, potentially leading to small signal-to-noise ratio when computing the Seebeck coefficient. Conversely, an accurate estimate of $S$ through the expectation value of the correlation coefficient, and $\sigma$ through the cepstral method, allows us to mitigate this issue.
Given the positive outcomes of the benchmarks on molten $\ce{CsF}$, we extend our analysis to compute the Seebeck coefficient from relatively short simulations on $\ce{KCl}$, $\ce{NaCl}$, and $\ce{LiCl}$ under a range of pressures and temperatures.

\subsubsection{Application to machine learning potentials}

\begin{figure}
    \centering
    \includegraphics[width=\linewidth]{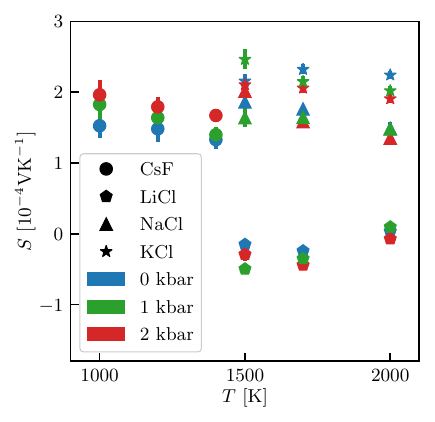}
    \caption{Seebeck coefficient of molten \ce{LiCl}, \ce{CsF}, \ce{NaCl} and \ce{KCl} as  functions of pressure and temperature. The Seebeck coefficient has been estimated via the Bayesian regression procedure as detailed in the text.}
    \label{fig:comparison_bayesian_materials}
\end{figure}

The \ce{NaCl}, \ce{KCl}, and \ce{LiCl} systems are simulated from a sample of $1000$ atoms interacting with a Deep Potential Molecular Dynamics (DeePMD) neural network potential~\cite{wang2018, zhang2018} obtained from Ref.~\cite{LIANG2021}. Trajectories are integrated with a $1\,\mathrm{fs}$ time step using the Velocity-Verlet algorithm as implemented in LAMMPS~\cite{LAMMPS}. We simulate different temperature-pressure conditions by equilibrating the systems in the $NpT$ ensemble for $500\,\mathrm{ps}$ and $NVT$ ensemble for $200 \,\mathrm{ps}$. After equilibration, the energy~\cite{tisi2021heat} and charge fluxes are harvested from a $2\,\mathrm{ns}$ canonical~\cite{bussi2007canonical} trajectory sampled every $2.5\,\mathrm{fs}$. Deep potentials do not feature explicit atomic charges. Nonetheless, integer oxidation states ($+e$ for cations, and $-e$ for anions, $e$ being the elementary charge) can be rigorously assigned to each atomic species to compute Onsager coefficients~\citep{grasselli2019topological,pegolo2020oxidation,pegolo2022topology}.

Fig.~\ref{fig:comparison_bayesian_materials} illustrates the Bayesian regression estimates of the Seebeck coefficient for molten \ce{CsF}, \ce{KCl}, \ce{LiCl}, and \ce{NaCl} as functions of pressure and temperature. While the experimental literature on the thermoelectric response of molten salts is limited, we can compare our results on molten \ce{NaCl} and \ce{KCl} with some experimental data on the thermopower, defined as the electric potential induced by a temperature difference~\cite{archer1963, kuan2021}. 
Our computed Seebeck coefficients show order-of-magnitude agreement with the experimental data in Ref.~\cite{archer1963}. However, it is challenging to draw conclusions on the agreement between simulations and experiments, whose differences might stem from factors such as the choice of electrodes in the experimental setup~\cite{bedeaux2023}, or the specific force field employed in the simulations.

Another relevant feature visible in Fig.~\ref{fig:comparison_bayesian_materials} is the different sign of $S$ in different molten salts. As already discussed by Onsager~\cite{onsager1939}, the mass discrepancy between ions can significantly influence the sign of the Seebeck coefficient. Notably, when the mass difference between cations and anions is substantial, as the cases of \ce{CsF} and \ce{LiCl}, a mass effect can be observed.  Our simulations corroborate this hypothesis, revealing a different sign reversal between \ce{CsF} and \ce{LiCl} due to their considerable cation/anion mass imbalance. 

\section{Conclusions}\label{sec:conclusions}

This work addresses the long-standing challenge of providing an accurate and statistically sound computation of the Seebeck coefficient in liquid electrolytes through equilibrium MD simulations. Leveraging the statistical properties of the heat-charge periodogram, we have devised an efficient and consistent Bayesian regression approach to compute the Seebeck coefficient from optimally short trajectories. We have benchmarked out methodology against extensive GK calculations on a simple semi-empirical model of molten \ce{CsF}. We then used the Bayesian approach to compute the Seebeck coefficients of \textit{ab initio} accurate models of molten \ce{LiCl}, \ce{NaCl}, and \ce{KCl} as a function of pressure and temperature. 

The Bayesian methodology not only enables a comprehensive characterization of the transport-coefficient phase diagram of electrolytes, but also paves the way to deeper investigations into ionic thermoelectric materials, as it enables the computation of all relevant transport coefficients.
The Bayesian results discussed in this work for the molten \ce{NaCl} and \ce{KCl} are also in fair agreement with some experimental measurements of the thermopower of molten alkali halides.\cite{archer1963}

\section*{Data Availability}
All the relevant data, scripts, and input files that support the results of this work are available on the Materials Cloud platform \cite{Materials-Cloud} at \url{https://doi.org/10.24435/materialscloud:p1-bm}.

\section*{Acknowledgments}
We are grateful to F. Grasselli for a critical reading of an early version of the manuscript. PP thanks M. Berti for insightful discussions regarding MCMC calculations.

This work was partially supported by the European Commission through the \textsc{MaX} Centre of Excellence for supercomputing applications (grant number 101093374) and by the Italian MUR, through the PRIN project ARES (grant number 2022W2BPCK) and the Italian National Centre for HPC, Big Data, and Quantum Computing (grant number CN00000013), funded through the \emph{Next generation EU} initiative.

\bibliography{main}

\end{document}